\documentclass[iop,letterpaper]{emulateapj}
\usepackage{graphicx}
\shorttitle{Probability Weighting for Periodicity Searches}
\shortauthors{Kerr}

\newcommand{\dg}{^{\circ}}
\newcommand{\fermi}{\emph{Fermi}-LAT }
\newcommand{\fermin}{\emph{Fermi}-LAT}
\newcommand{\phis}{$\{\phi_i\}$ }
\newcommand{\vom}{\vec{\Omega}}
\newcommand{\vla}{\vec{\lambda}}
\newcommand{\mcf}{\mathcal{F}}

\newcommand{\fluxunits}{ph cm$^{-2}$ s$^{-1}$ }
\newcommand{\fluxunitsn}{ph cm$^{-2}$ s$^{-1}$}
\newcommand{\zt}{$Z^2_m$}

\begin{document}

\title{Improving Sensitivity to Weak Pulsations with Photon Probability Weighting}
    
\author{M. Kerr\altaffilmark{1}}
\affil{Kavli Institute for Particle Astrophysics and Cosmology, Stanford University, 452 Lomita Mall, Stanford, CA 94305, USA}
\email{kerrm@stanford.edu}
\altaffiltext{1}{Einstein Fellow}

\begin{abstract}
All $\gamma$-ray telescopes suffer from source confusion due to their inability
to focus incident high-energy radiation, and the resulting background
contamination can obscure the periodic emission from faint pulsars.  In the
context of the \emph{Fermi} Large Area Telescope, we outline enhanced
statistical tests for pulsation in which each photon is weighted by its
probability to have originated from the candidate pulsar.  The probabilities
are calculated using the instrument response function and a full spectral
model, enabling powerful background rejection.  With Monte Carlo methods, we
demonstrate that the new tests increase the sensitivity to pulsars by more than
$50\%$ under a wide range of conditions.  This improvement may appreciably
increase the completeness of the sample of radio-loud $\gamma$-ray pulsars.
Finally, we derive the asymptotic null distribution for the $H$-test, expanding
its domain of validity to arbitrarily complex light curves.
\end{abstract}


\section{Introduction}
Though the detection and characterization of periodic emission from neutron
stars has historically been the province of radio astronomers
\citep[e.g.][]{hewish68,msp_discovery,parkes_multibeam_1}, increasingly sensitive
instruments have enabled the study of pulsars at high energy.  Neutron stars
accreting near their Eddington limit were detected in X-rays as
accretion-powered pulsars \citep{xray_apps} in the 1970s by UHURU
\citep{giacconi_et_al1971}, while the sensitive ROSAT and ASCA missions detected
faint magnetospheric emission from a population of rotation-powered X-ray
pulsars \citep{xray_rpps}.

At even higher energies, the handful of well-known $\gamma$-ray pulsars---e.g.,
Vela \citep{sas2_vela} and Geminga \citep{egret_geminga_detection}---have been
joined by a host of new pulsars detected by the \emph{Fermi} Large Area
Telescope (\fermin).  While the superb sensitivity of \fermi has facilitated 
the first discoveries of pulsars in $\gamma$ rays alone
\citep{cta1,blind_search_16}, a foundation of pulsar science with \fermi is the
extensive support from the radio community.  E.g., the Pulsar Timing Consortium
\citep{smith_timing} generates timing solutions for over 200 pulsars with high
spindown luminosity, $\dot{E}>10^{34}$ erg s$^{-1}$.  These timing solutions
enable the long integrations necessary to detect periodicity in sparse \fermi
photons, and so far more than 30 timed pulsars have been detected in $\gamma$
rays \cite[e.g.]{fermi_j2021p3651,psr_cat,msp_pop}.

Although the precise emission geometry of pulsars is still unknown, it is not
unreasonable to believe that $\approx50\%$ of radio-loud pulsars are also
$\gamma$-ray loud \citep{ravi_etal_2010}, a fraction that may be even larger
for millisecond pulsars.  If this is indeed the case, then many of the luminous
radio-timed pulsars are visible in $\gamma$ rays but below the current
sensitivity of the LAT.  While \fermi continuously observes the GeV sky, the
flux above which pulsars are detectable only decreases as $t^{-1/2}$. Ten years
of observation will only decrease the current flux threshold by a factor of
$\sim$2.

Improved analysis techniques can beat this rate, and in this vein we outline
better statistics for testing for periodicity.  To date, such tests
\citep{psr_cat} have used only the arrival time/phase of a photon.  As shown by
\citet[][hereafter BKR08]{bickel_kleijn_rice_2008}, the additional data
available---the photon's reconstructed energy and position---allow the
calculation of a probability that the given photon originated from the
candidate pulsar, and that incorporating this probability into the test
statistics helps reject background and increase the sensitivity to pulsations.
Although we focus on the application of the technique to LAT data, we note that
the scheme is applicable to \emph{any} photon-counting instrument in which
sources are not perfectly separated from their background, e.g. searches for
X-ray pulsation in observations of a pulsar embedded in a pulsar wind nebula.

The paper is organized as follows.  We begin in Section \ref{sec:stats} by
giving an overview of a family of statistics---based on trigonometric moments
and formulated as a score test for pulsation---that includes the weighted
pulsation tests discussed here. In Sections \ref{sec:z2m}, we review the \zt
and $H$-test statistics and define modified versions incorporating weights.  We
outline the calculation of probability weights appropriate for the LAT in
Section \ref{sec:prob_weights} and in Section \ref{sec:performance} we
demonstrate the superior performance of the weighted versions of the \zt and
$H$-test statistics.  In the Appendix, we derive the asymptotic calibration for
the $H$-test, a new result expanding the scope of the test.

\section{Statistical Tests for Periodicity}
\label{sec:stats}

A timing solution (ephemeris) defines a map from photon arrival time $t$ to phase
$\phi$, e.g. neutron star rotational phase.  The flux from a pulsar can be written as
\begin{equation}f(\phi,E) \propto 1 + \eta\,\sum\limits_{k=1}^{\infty}
\alpha_k \cos(2\pi k\phi) + \beta_k \sin(2\pi k\phi).
\end{equation}
For simplicity, we have assumed the pulsar spectrum is independent of phase.

The null hypothesis---no pulsation---is given by $\eta=0$.  By considering the
likelihood for photon arrival times, BKR08 derived a test statistic\footnote{The
form of the statistic presented here requires that the pulsar period be short
relative to the timescale on which the detector response changes.  This is so
for the \fermin.} for $\eta>0$,
\begin{equation}
\label{eq:q}
Q \equiv \frac{2}{T} \sum_{k=1}^{\infty} (\alpha_k\,\hat{\alpha}_k)^2 +
(\beta_k\,\hat{\beta}_k)^2,
\end{equation}
where $T$ is the total integration time and
\begin{equation}
\label{empir_trig}
\hat{\alpha}_k = \sum_{i=1}^{n} w_i\,\cos(2\pi k\phi_i);\ \
\hat{\beta}_k = \sum_{i=1}^{n} w_i\,\sin(2\pi k\phi_i),
\end{equation}
where the sum is over the list of $n$ photons and $w_i$ is some weight.  (See
Eq. 17 of BKR08, from which this form is adapted.)  For $w_i=1/n$,
these are estimators of the trigonometric moments of the distribution and Monte
Carlo estimators for the coefficients of the Fourier transform.  If $w_i$ is
the probability that a photon comes from the pulsar, the weights are
optimal in the sense that $Q$ is a score test, which is locally most
powerful.  Moreover, the statistic is manifestly invariant under phase shifts
$\phi\rightarrow\phi+\delta\phi$.  (See also the result of \citet{beran1}, who
considered a similar class of statistics.) From Central Limit Theorem
arguments, BKR08 show that for a finite collection of $m$ harmonics, $Q$
has an asymptotic distribution of $\chi^2_{2m}$.

\subsection{The $Z^2_m$ Test}
\label{sec:z2m}
A simple realization of such a statistic, with $w_i=1$ and $\alpha_k=\beta_k=1$
for $k\leq m$ and $\alpha_k=\beta_k=0$ for $k > m$, known as the \zt
statistic,
\begin{equation}
\label{eqn:ztest}
Z^2_m = \frac{2}{n} \sum_{k=1}^{m} \hat{\alpha}_k^2 + \hat{\beta}_k^2,
\end{equation}
has been a workhorse for searches for $\gamma$-ray pulsars.  (Note the
change in normalization: $1/T\approx1/n$.)  A $Z^2_2$ test was used in a search
for pulsations in COS-B data using timing solutions for $145$ radio pulsars
\citep{cos-b_pulse_search}.  A similar search of EGRET data
\citep{egret_pulse_search} used $Z^2_m$ tests with 1, 2, and 10 harmonics, the
$H$-test (see below), and the ``$Z^2_{2+4}$'' test which is defined as above
but with summation restricted to the 2nd and 4th harmonics.  $Z^2_m$ forms an
integral part of the $H$-test, and continues to see use in analysis of \fermi
data \citep{psr_cat}.

From the discussion above, we expect $Z^2_m$ to be distributed as $\chi^2_{2m}$
in the null case, but this result holds if and only if the $\hat{\alpha}_k$ and
$\hat{\beta}_k$ coefficients are statistically independent.  However, in some
cases they may be highly dependent. For a single observation, $\alpha_k$
($\beta_k$) can be inverted to find the original rv, $\phi$, and thus all
coefficients are algebraically determined.  For large samples from a uniform
distribution, however, a given coefficient conveys little information about the
underlying \phis and the coefficients are approximately independent.  Although
the requirement on the detector response assures uniformity, if the null
distribution is peaked, the coefficients remain correlated for arbitrarily
large sample sizes.

\begin{figure*}
\center
\includegraphics[width=6.5in]{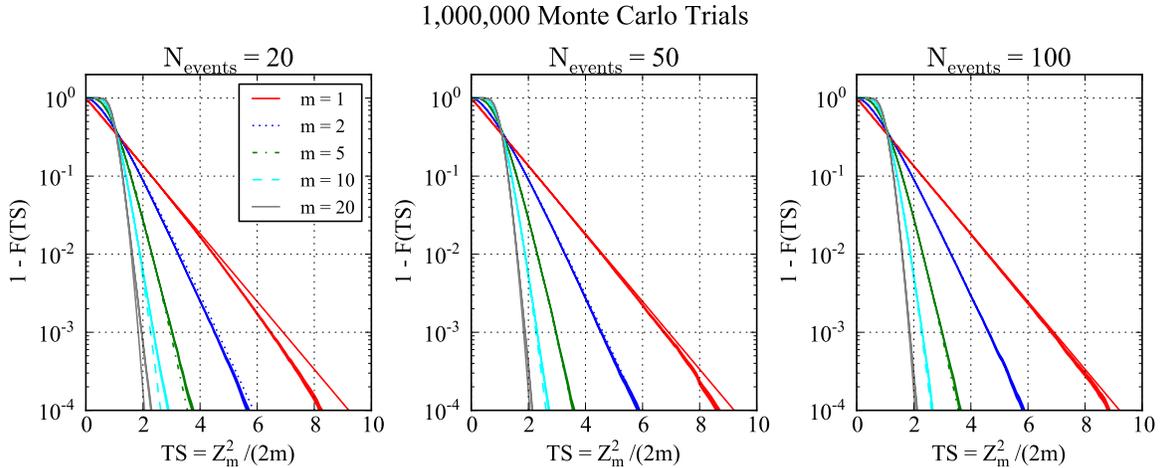}
\caption{The observed distribution for the $Z^2_m$ statistic for varying sample
size and maximum harmonic ($m$).  The asymptotic calibration is shown as a solid
line, while the solid band gives the empirical distribution function of the
Monte Carlo realizations.  The width indicates the statistical uncertainty
estimated as $\sqrt{N(>=TS)}$.
}
\label{ch5_plot1}
\end{figure*}

To determine the minimum sample size required to reach the asymptotic $\chi^2$
distribution, we performed a Monte Carlo study of the convergence as a function
of both sample size and maximum harmonic ($m$), shown in Figure
\ref{ch5_plot1}.  Evidently, a sample size of at least $50$ phases is required
for robust significance estimation at the $3\sigma$ level, and convergence
appears to improve for higher values of $m$.

The defintion of the \zt can be viewed as a set of hard cuts for which
$w_i=1$ for some set of photons and $w_i=0$ for all other photons.  By allowing
arbitrary values for $w_i$ we implement soft cuts in which we ideally assign
higher (lower) weights to photon associated with the pulsar (background).  We
thus define the \emph{weighted} \zt as 
\begin{equation}
\label{eq:weighted_z2m}
Z^2_{2mw} \equiv
\frac{2}{n}\,\left(\frac{1}{n}\sum\limits_{i=1}^{n}{w_i^2}\right)^{-1}
\sum_{k=1}^{m} \hat{\alpha}_{k}^2 + \hat{\beta}_{k}^2.
\end{equation}
As a realization of the test statistic of Eq. \ref{eq:q}, its
calibration remains $\chi^2_{2m}$.  Again, note the expression of the
normalization in terms of the data; the relation can be seen in terms of a
random walk.

\subsection{ The $H$-Test }
\label{sec:htest}

It is clear from Eq. \ref{eq:q} that for an optimal test, the empirical
coefficients should be weighted by the true Fourier coefficients, which are
\textit{a priori} unknown.  The estimation of unit coefficients up to a maximum
harmonic $m$ in the \zt is crude, and choosing $m$ too small will result in a
loss of power against sharply-peaked light curves, while $m$ too large will
lose power against broad, sinusoidal light curves.  To improve on this,
\citet{dejager_1} proposed the $H$-test, which estimates $m$ from the data.
They defined
\begin{equation}
\label{eqn:htest}
H_m = \max\left[Z^2_i - c\times(i-1)\right], 1 \leq i \leq m
\end{equation}
and specifically recommended $m=20$ and $c=4$ as an omnibus test.  They
provided a Monte Carlo calibration of the tail probability and, in a recent
paper \citep{dejager_2} provide an estimate
$1-F_{H_{20}}(h)\approx\exp(-0.4\times h)$ good to $h\approx70$.  We derive the
analytic, asymptotic calibration for all values of $m$, $c$, and $h$ in the
Appendix and use this calibration anywhere a conversion from $h$ to chance of
Type I error (i.e., ``$\sigma$'') is needed.

This calibration depends only on the asymptotic
$\chi^2_{2m}$ calibration of $Z^2_m$, and so it also holds for $Z^2_{mw}$ and
thus for a \emph{weighted H-test statistic}
\begin{equation}
H_{mw} \equiv \max\left[Z^2_{iw} - c\times(i-1)\right], 1 \leq i \leq m.
\end{equation}
We adopt the original values $m=20$ and $c=4$ below.

\section{Calculating Photon Probabilities}
\label{sec:prob_weights}

The optimal choice of weight is the \emph{probability} that a photon originates
from the pulsar, and we outline the calculation of this quantity for the
\fermin.  We have available three pieces of information, viz. the
photon's reconstructed energy, position, and arrival time.  For pulsars, the
arrival time is necessary for computation of the phase but can otherwise be
ignored except in rare cases, e.g.  a candidate near a bright, highly variable
blazar.  We thus consider \emph{time-averaged} quantities only.  (Recall we
have assumed phase-independent pulsar spectra.  Spectra generally \emph{are}
dependent on phase \citep[e.g.][]{vela2}, but such using a phase-averaged
spectrum incurs little error.)

Probabilities constructed using only the photon position have been employed
profitably for COS-B and EGRET analyses \citep{clemson,ramanamurthy,maura}.
This approach has the advantage of requiring no knowledge of source spectra and
works well for a detector with an excellent and/or energy-independent psf.  The
\fermin, on the other hand, has a comparatively poor psf that varies by more
than two orders of magnitude between $100$ MeV and $100$ GeV
\citep{lat_instrument}.  A typical source will have many background photons
within the psf radius at low energy but very few at high energies, and a
probability based solely on position cannot discriminate against the many low
energy background photons.

We address the strong dependence of S/N on energy by including the estimated
photon energy in the probability calculation.  Briefly, a point source is
characterized by its photon flux density (ph cm$^{-2}$ s$^{-1}$ MeV$^{-1}$)
which we model as $\mcf(E,\vla)$, with $E$ the energy and $\vla$ some set of
parameters (e.g., the normalization and photon index for a power law; these
parameters will typically be estimated via maximum likelihood analysis as
discussed in the following section).  We assume the source is stationary.  If
the exposure\footnote{The exposure calculation for the \fermi involves summing
the detector response over the pointing history of the spacecraft.} to the
position of the source, $\vom_0$, is given by $\epsilon(E,\vom_0)$ (cm$^{2}$
s), then the expected differential rate (ph MeV$^{-1}$ sr$^{-1}$) in the
detector from the $j$th source is \begin{equation} r_j(E,\vom) = \mcf(E,\vla)\,
\epsilon(E,\vom_0)\, f_{\mathrm{psf}}(\vom ; \vom_0, E), \end{equation} where
$f_{\mathrm{psf}}(\vom ; \vom_0, E)$ denotes the psf of the instrument for the
incident energy and position.  While the psf only depends on incidence angle,
the highly structured diffuse background motivates preservation of the full
position.

For a photon observed with energy near $E$ at a
position near $\vom$, the probability that it originated at the $j$th source is
simply
\begin{equation}
\label{eq:weights}
w_j (E,\vom|\vla) \equiv \frac{r_j(E,\vom,\vla_j)}{\sum\limits_{i=1}^{N_s} r_i(E,\vom,\vla_i)},
\end{equation}
where $N_s$ is the number of contributing sources.  The weight can approach $1$
for bright pulsars and at high energies, where the LAT psf becomes narrow.

\section{Performance in \emph{Fermi}-LAT Pulsation Searches}
\label{sec:performance}

The primary goal in adopting the weighted versions of pulsation test statistics
is, of course, to find more genuine pulsars.  In statistical language, we want
to minimize Type I Error (false positives) and Type II error (false negatives).
We address these two in turn by comparing weighted and unweighted versions of
the \zt and H-test statistics.  Although BKR08 provide an analytic expression
for the increased detection significance offered by photon weights (see Eq. 23
of that work), it is not suitable for determining the global performance
improvement since the optimal data selection for the unweighted tests is
unknown \textit{a priori}. We thus assess the performance of the tests on an
ensemble of simulated pulsars.

\subsection{Simulation Details}
The \fermi Science
Tool\footnote{http://fermi.gsfc.nasa.gov/ssc/data/analysis/scitools/overview.html}
\emph{gtobssim} uses a detailed characterization of the instrument response
function to simulate events from modeled point and diffuse sources.  We
simulate a point source---the candidate pulsar---with a realistic pulsar
spectrum,
\begin{equation}
\label{eq:vela_spec}
\frac{dN}{dE} \propto (E/\mathrm{GeV})^{-\Gamma} \exp( -E/E_c)
\end{equation}
with $\Gamma=1.5$ and $E_c = 3$ GeV.  This spectrum is typical of many pulsars
and is approximately that of the middle-aged Vela pulsar \citep{vela1}.  We
place the source at (R.A., Decl.) = (128.8463, -45.1735), the position of Vela.
For the background, we simulate photons from the two diffuse background models
used in the 1FGL catalog analysis \citep{1fgl}, {\it gll\_iem\_v02}---a model of
the Galactic diffuse background due to cosmic rays---and {\it
isotropic\_iem\_v02}---an isotropic background including contributions from
unresolved extragalactic point sources and instrumental backgrounds.  In all, we
simulate one year of integration using the spacecraft pointing history from
2009.

The normalization of Eq. \ref{eq:vela_spec} is chosen to yield a bright source
with an integral photon flux from 100 MeV to 100 GeV of $\mcf_{sim}\equiv
10^{-5}$ \fluxunitsn.  We are interested in detection of dim sources, so from
this set of photons we select subsets emulating sources with
appropriately lower fluxes, e.g. $\mcf_{tar} = 10^{-8}$ \fluxunitsn, by (a)
drawing the target number, $N_{tar}$, of photons from a Poisson distribution
with mean $N_{tot}\times\mcf_{tar}/\mcf_{sim}$ and (b) selecting a subset of
$N_{tar}$ of the original photon set at random (without replacement).  In this
way, we can generate ensembles of statistically independent point sources over
a range of fluxes from a single Monte Carlo data set.  While we use a single
realization of the diffuse photons, we effectively generate a new iteration by
randomizing these photons in phase for each ensemble member as we discuss
below.

To determine the weights (Eq. \ref{eq:weights}), we employ \emph{gtsrcprob}.
This Science Tool combines the source models (those used to generate the Monte
Carlo data) with the instrument response function to determine the observed
source rates and hence the weights.  These weights are valid for the simulated
point source flux $\mcf_5$, and must be scaled for the dim ensemble members:
$w_{tar}^{-1} - 1 = (w_{sim}^{-1} - 1)\times\mcf_{sim}/\mcf_{tar}$.

The Monte Carlo events as generated by \emph{gtobssim} have no pulsation.
During simulation, each photon is ``tagged'' with an identifier for its
originating source.  We assign phases from a uniform distribution to the
diffuse background, and for the point source we draw phases from an assumed
light curve, typically a normalized sum of wrapped Gaussians.

\subsection{Performance: Type 1 Error}
\label{subsec:type1}

Type I error stems from two sources.  First, there is the chance of a
fluctuation in the test statistic (TS) sufficiently large to pass the
established threshold for rejection of the null hypothesis, i.e., claiming
detection of a pulsar.  As long as we understand the null distribution of the
TS, this particular source of error is easy to control: we simply determine in
advance our tolerance to false positives and set the TS threshold accordingly.
We must be cautious about applying the asymptotic calibration of the null
distribution to small sample sizes.  For these cases, it is important
to verify the chance probability with a Monte Carlo simulation.

A second, more insidious source of error arises from the strong influence of
the data selection scheme on the pulsed S/N, this dependence stemming from
the energy-dependent psf, source confusion, the strong Galactic diffuse, and
the exponential suppression of pulsar emission above a few GeV.  To find the
best cuts, one is tempted to use the TS itself as a metric, a procedure which
invalidates its calibration.  Failure to account for this change increases the
probability of false positives.  Stringent cuts may also make the asymptotic
calibration poorer.

Probability-weighted statistics eliminate these problems.  Since the weights
naturally go to zero for photons with neglible signal, one could in principle
include \emph{all} LAT data in the TS.  (In practice, little signal is
contained in photons more than $2^{\circ}$ from the source.) Weighting provides
an amorphous, optimal selection reflecting, e.g., the proximity of the Galactic
plane or the estimated pulsar cutoff energy\footnote{In this regard, the
weights deliver a data set similar to that proposed by \citet{hans}, who
developed an algorithm for determining an optimal aperture with an arbitrary
``edge'' determined by the local signal-to-noise ratio}. And this single
selection incurs no probability of Type I error beyond that due to statistical
fluctuations.    Finally, the weighted statistic is less susceptible to
small-sample effects since all relevant information is included, though for
particularly weak signals Monte Carlo validation remains important. 

\begin{figure}
\center
\includegraphics[width=3.5in]{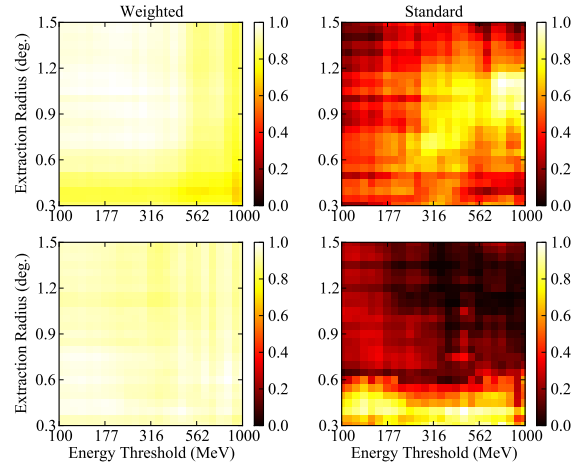}
\caption{A comparison of the dependence on event selection of the $H_{20w}$ and
$H_{20}$ statistics.  Two members of an ensemble (generated as described in the
main text) with flux $10^{-8}$ \fluxunits and a single-peaked light curve of
Gaussian shape ($\sigma=0.03$) are shown, one in each row.  The two test
statistics were calculated over a grid of data selection criteria: the y-axis
gives the extraction radius (maximum angular distance from the pulsar) and the
x-axis indicates the threshold energy below which photons are excluded.  The
lefthand (righthand) columns shows the results for the weighted (un-weighted)
test statistic after conversion to $\sigma$ units and normalization to the
maximum observed value.} \label{ch5_plot8}
\end{figure}

To make these claims concrete, we compare the weighted $H$-test ($H_{20w}$) and
the standard $H$-test ($H_{20}$) computed over a grid of cuts on photon
position and energy (Figure \ref{ch5_plot8}).  The unweighted statistics show
strong TS peaks for certain energy thresholds and extraction radii, and these
peaks vary from realization to realization, precluding an \textit{a priori}
calculation of an optimal aperture.  The weighted statistics, on the other
hand, are largely insensitive to the data selection and perform best for a
simple prescription: use as many photons as is practical. We are thus free to
use the same loose cuts for all sources, maintaining good performance (peak TS)
without the need to tune.

\subsection{Performance: Type II Errors / Sensitivity}

We must also consider Type II error---false negatives.  In astrophysical terms,
this error rate is essentially the sensitivity of the method, i.e., the flux
threshold above which a typical source will be significantly detected in a given
observation.  Indeed, we adopt this approach to \emph{define} the Type II error.
Given a particular light curve, we generate ensembles of sources at a series of
increasing fluxes until we identify the flux at which a given fraction of the
ensemble has a detection significance above threshold.

More specifically, to compute the detection significance, we calculate the tail
probability of the asymptotic distribution for the TS in question convert it to
(two-tailed) $\sigma$ units, i.e., a chance probability of $X\sigma$ is $1 -
(2\pi)^{-1/2}\int_{-X}^{X}dx\,\exp(-x^2/2)$.  We determine the flux threshold
as that for which $68\%$ of the ensemble deliver a $\sigma$ value above a
pre-determined threshold (see below).  This is the \emph{detection flux
threshold}.  To estimate the $68\%$ level robustly, we fit the ensemble values
with a normal distribution and report the appropriate quantile.

\begin{figure}
\center
\includegraphics[width=3.5in]{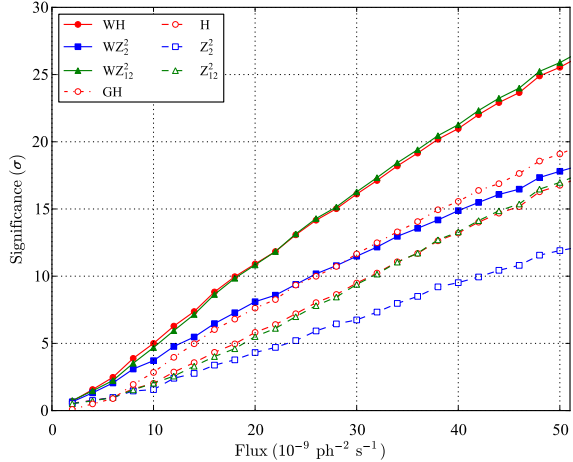}
\caption{The dependence of significance in $\sigma$ units as a function of flux.
The reported value at a given flux is that attained by at least $68\%$ of the
ensemble of 50 MC realizations.  The trend is approximately linear, as explained
in the main text, making it simple to invert the relation.  The light curve here
is a single Gaussian peak with $\sigma=0.03$.}
\label{ch5_plot10}
\end{figure}

To determine the flux threshold, we need to invert the calculated quantity
(tail probability in $\sigma$ units) to the desired quantity (flux threshold).
Surprisingly, the relationship between the two is linear.  Significance
canonically scales as $\sqrt{n}$, or in this case, since we integrate for
exactly one year, $\sqrt{\mcf}$.  However, here we increase the source flux
without increasing the background flux, so we also increase the S/N, and in the
background-dominated r\'{e}gime, significance is proportional to the S/N.
Thus, $\sigma\propto\mcf$, demonstrated in Figure \ref{ch5_plot10}.

We choose $4\sigma$ as the threshold which causes us to discard the null
hypothesis, i.e., to claim detection of periodic emission.  This threshold
corresponds to a very small probability of Type I error, about 1 in 15800.
Selecting an even higher threshold---and consequently applying the asymptotic
calibration well into the distribution's tails---is not appropriate for these
small sample sizes.

To give a modest survey of the statistics outlined above, we present weighted
and un-weighted versions of $H_{20}$, $Z^2_{12}$, and $Z^2_2$.  Recall that
$H_{20}$ is an omnibus test that depends little on light curve morphology,
whereas $Z^2_{12}$ ($Z^2_{2}$) should perform well for light curves with sharp
(broad) features.  For the unweighted versions of these tests, we select
photons with $E > 200$ MeV and an angular separation from the candidate
$\leq0.8\dg$.  However, since no single extraction criterion will yield optimal
values for the unweighted statistics, we also include a ``grid search'' for
$H_{20}$.  That is, for each source, we extract photons with $E>E_{th}$ and
$\theta < \theta_{th}$, i.e., reconstructed energies above $E_{th}$ and
reconstructed positions separated from the true position by less than
$\theta_{th}$.  We do this for a grid of $E_{th}$ and $\theta_{th}$ with
$E_{th}/\mathrm{MeV}\in(100,178,316,562,1000)$ and
$\theta_{th}/\mathrm{deg.}\in(0.5,0.625,0.75,0.875,1.0)$.  We take the maximum
resulting test statistic, convert it to chance probability, multiply by 25 (the
number of ``trials''), and convert this quantity to $\sigma$ units.  Finally,
for the weighted statistics, we select photons with $E > 100$ MeV and $\theta
\leq2\dg$.

\begin{figure}
\center
\includegraphics[width=3.5in]{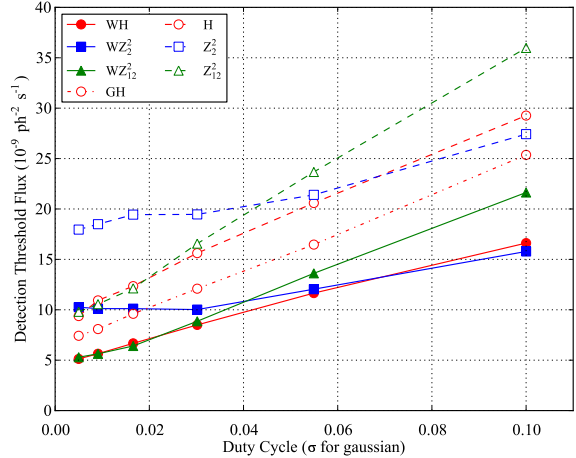}
\caption{The $68\%$ flux detection threshold based on a $4\sigma$ detection
criterion for a single-peaked Gaussian light curve as a function of duty cycle.
Light curves with narrow (broad) peaks lie to the left (right).  $H_{20}$,
$Z^2_{12}$, and $Z^2_2$ are the unweighted statistics with a single data
extraction, and $WH_{20}$, $WZ^2_{12}$, and $WZ^2_2$ are the weighted versions
of these statistics.  $GH$ is the gridded, unweighted $H_{20}$ statistic.}
\label{ch5_plot4}
\end{figure}

\begin{figure}
\includegraphics[width=3.5in]{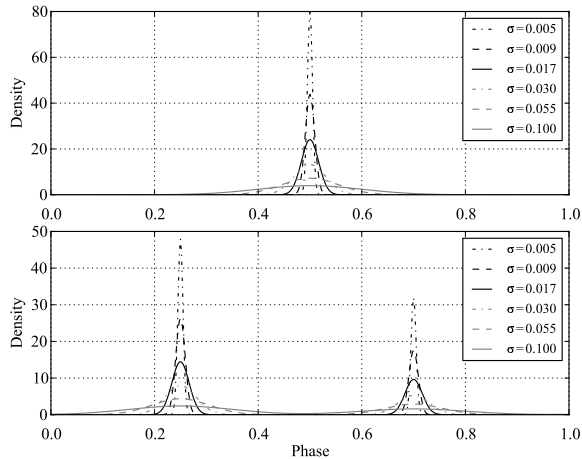}
\caption{The templates used for the single- and double-peaked light curves in
the determination of the detection flux thresholds.  The functional form is a
wrapped Gaussian, $f(\phi)= \sum_{i=-\infty}^{\infty}\,g(\phi + i)$ with
$g(\phi,\mu,\sigma) = (2\pi)^{-0.5}\,\exp(-0.5\,(\phi-\mu)^2/\sigma^2)$.  In the
first panel, $\mu=0.5$, while in the second panel, $\mu_1=0.25$, and
$\mu_2=0.70$.  In this panel, the ratio of the peak heights is 3/2, and the peak
widths are identical.}
\label{ch5_plot11}
\end{figure}

The primary result, shown in Figure \ref{ch5_plot4}, shows the flux threshold
for each method as a function of ``duty cycle'', the fraction of the full phase
for which there is appreciable pulsed emission.  In this case, the light curves
are single Gaussian peaks with a variety of values for their $\sigma$ (standard
deviation) parameter; the templates are shown in Figure \ref{ch5_plot11}.  For a
fixed flux, increasing the duty cycle decreases the peak flux, or S/N, and the
primary dependence is then an inverse relation between flux threshold and duty
cycle.  However, there is additional dependence from the nature of each test.
It is clear, e.g., that the $H_{20}$ and $Z^2_{12}$ perform significantly better
than $Z^2_2$ for low duty cycle sources, while $Z^2_2$ maintains a slight edge
for broad light curves.  The $H$-test does a good job for all duty cycles.  More
importantly, it is clear that the weighted statistics enjoy a lower threshold
for detection---by a factor of 1.5 (grid search) to 2.0 (single cut)--- than
their unweighted counterparts.

\subsubsection{Two-peaked light curves}
\begin{figure}
\includegraphics[width=3.5in]{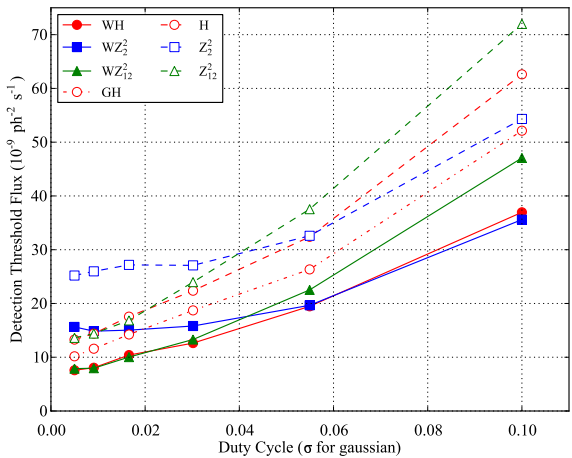}
\caption{As Figure \ref{ch5_plot4}, but for a double-peaked Gaussian light curve.}
\label{ch5_plot12}
\end{figure}

Many pulsar light curves display two peaks, often separated by about 0.4-0.5
cycles \citep{psr_cat}.  We repeat the analysis of the previous section using a
template comprising two Gaussian peaks separated by $0.45$ in phase.  Real
pulsar peaks often have unequal intensities (with an energy-dependent ratio.)
We reflect that here with a slightly-dominant leading peak; see Figure
\ref{ch5_plot11}.  As seen in Figure \ref{ch5_plot12}, the overall flux
thresholds are unsurprisingly increased: at a fixed flux, spreading photons
between multiple peaks decreases the overall S/N.  The weighted statistics
maintain a comfortably decreased flux threshold with respect to the unweighted
methods.

\subsubsection{Pulsars with DC Emission Components}

\begin{figure}
\includegraphics[width=3.5in]{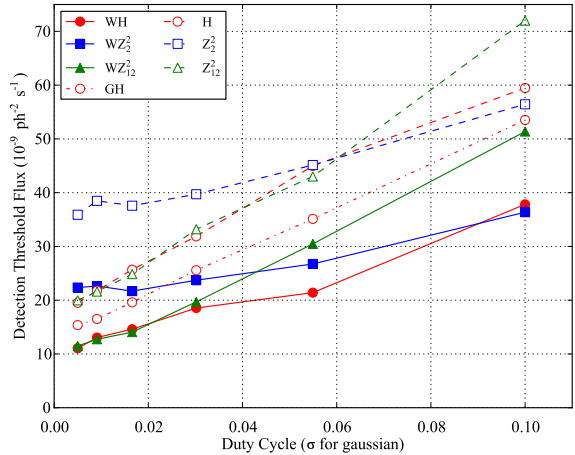}
\caption{The $68\%$ flux detection threshold based on a $4\sigma$ detection
criterion for a single-peaked Gaussian light curve as a function of duty cycle.
Here, the pulsed fraction has been decreased to $50\%$, i.e., half of the
photons from the source are uniformly distributed and half are drawn from the
Gaussian light curve.  The flux threshold for the weighted statistic is
$1.5-2.0$ times lower than the unweighted statistics, with some slight
dependence on duty cycle and statistic.}
\label{ch5_plot13}
\end{figure}

Some pulsars emit an appreciable flux of \emph{unpulsed} $\gamma$ rays, e.g. PSR
J1836+5925 \citep{j1836p5925}.  This steady emission is a confounding factor for
the weighted tests, since unpulsed source photons receive the same high
probability weights as pulsed photons but are distributed uniformly in phase,
decreasing the effective S/N.  However, for modest ratios of pulsed to unpulsed
flux, the performance of the weighted test is not unduly diminished.  In Figure
\ref{ch5_plot13}, we have examined the flux threshold for a single-peaked light
curve with equal contributions of pulsed and unpulsed emission.  As expected, we
see an overall increased flux threshold of about 2 over the fully-pulsed source
(c.f. Figure \ref{ch5_plot4}).  Despite a slight increase in the ratio of
weighted-to-unweighted thresholds, the weighted statistics still offer
appreciably improved sensitivity.  

\subsubsection{Effect of Uncertainties in Spectral Parameters}
\label{ch5:uncertain_spectrum}
From these demonstrations, it is clear that the probability-weighted statistics
have, on average, a factor of $1.5$ to $2$ improved sensitivity relative to the
unweighted versions.  One potential objection is that we have used the known
spectra in determining the photon weights, whereas with real data, of course,
we must first estimate the spectra of source and background.  To assess the
impact of using estimated parameters with concomitant uncertainty, we calculate
the weights based on parameters estimated via maximum likelihood spectral
analysis with the \textit{pointlike} tool \citep{my_thesis}.

For this test, we simulated 20 realizations of the Vela-like point source.
From Figure \ref{ch5_plot4}, we see that the detection threshold is about
$\approx8\times10^{-9}$\fluxunits, and we select this for the ensemble flux.
This flux yields about 100 source photons for the 1-year integration
period, allowing for asymptotic calibration.  Since the background strongly
affects the spectral analysis, independent realizations of the diffuse
background are important and accordingly we also simulated 20 realizations of
the diffuse sources.  To the point source data we added phase from a
single-Gaussian light curve with $\sigma=0.03$, while we generated uniform
random phases for the diffuse photons.

To assess the impact of uncertain parameters, we first calculate the
probability weights using the known model parameters for the pulsar and the
diffuse background, i.e., the ``ideal'' case.  We then perform a maximum
likelihood spectral fit to estimate the spectral parameters.  Since the
simulated point source is very dim relative to the background, it is impossible
to fit all three parameters.  We therefore fix the cutoff energy to $100$ GeV,
effectively a power law spectrum.  This approach is \emph{conservative} since
we are now using an incorrect spectral model.  Using the best-fit values for
the flux density and the photon index, we calculate a new set of probability
weights.  Finally, we compute $H_{20w}$ to determine the significance (a) using
the ``ideal'' weights and (b) using the ``measured'' weights.

\begin{figure}
\includegraphics[width=3.5in]{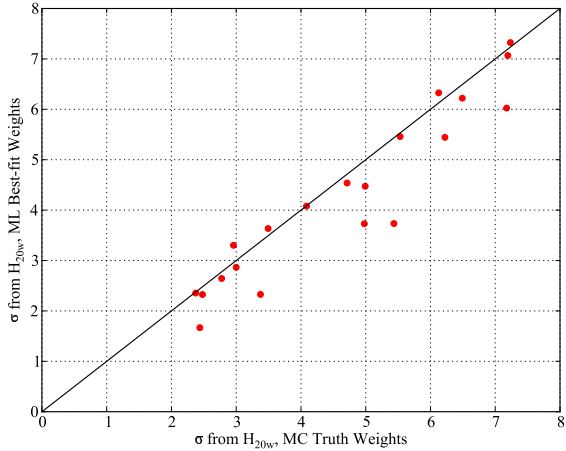}
\caption{The significance for each member of the ensemble described in the text
calculated using weights derived from the model used to simulated the data (the
Monte Carlo ``truth'') and derived from the ML model.  The ML-derived values are
very similar to those obtained using the known spectrum.}
\label{ch5_plot14}
\end{figure}

We compare the results for the two cases---in $\sigma$ units---in Figure
\ref{ch5_plot14}.  In general, the detection significance obtained with the
``measured'' weights is slightly lower than that obtained with the ``ideal''
weights, although in a few cases statistical fluctuations lead to the opposite
outcome.  Comparing the population means, the overall significance using
measured weights is decreased to $92\%$ of the ideal case, corresponding to an
increase in flux threshold of $\sim$10\%.  This effect is small compared to the
factor of $1.5-2.0$ increase in flux threshold seen between the weighted and
unweighted versions of the $H$-test.  We therefore conclude that---even
accounting for uncertainties in the spectral parameters used to calculate the
probability weights---weighted statistics offer a significant improvement in
sensitivity.

\subsubsection{Comparison of Pulsed and Unpulsed Detection Thresholds}
The machinery established above---performing spectral fits on an ensemble to
compute the weighted statistics using probabilities estimated from an ML
fit---also provides for directly comparing the DC (unpulsed) source significance
with the pulsed significance.  To estimate the DC significance, we use the
result of \citet{chernoff_miracle} for the asymptotic calibration for the
likelihood ratio for a single parameter whose null value lies on a boundary.  To
apply this calibration, we take the free parameter to be the flux density (which
has a boundary at 0) and fix the photon index $\Gamma=2.0$.  The cutoff remains
fixed at 100 GeV.  With this convention, the DC significance is then given by
\begin{equation}
\sigma_{DC} = \sqrt{2\times\log\mathcal{L}_{opt}/\mathcal{L}_0},
\end{equation}
with $\mathcal{L}_{opt}$ the likelihood value obtained with the best-fit flux
density and $\mathcal{L}_0$ the likelihood value obtained with the flux density
set to zero.  It should be noted that $\sigma_{DC}$ is a \emph{one-sided}
significance, so the same value of $\sigma_{DC}$ corresponds to a slightly
higher value the ``$\sigma$'' for the pulsed detection.  This discrepancy is
small compared to the observed difference in significance.

As in the previous section, we calculate the probability weights using both the
Monte Carlo truth values of the parameter and with the best-fit spectrum---in
this case, with cutoff energy \emph{and} photon index fixed---to estimate a
pulsed significance with the $H_{20w}$ test.

\begin{figure}
\includegraphics[width=3.5in]{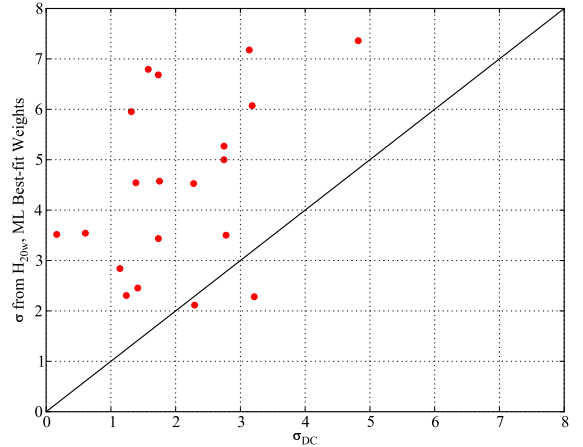}
\caption{A comparison of the significance of pulsed detection versus unpulsed
detection.  The pulsed detection significance was calculated using the ML
best-fit spectrum with the weighted $H$-test, while the unpulsed significance
was derived from a likelihood ratio test as described in the text.  For nearly
all sources, the pulsations are more strongly detected than the DC emission.}
\label{ch5_plot15}
\end{figure}

We compare the unpulsed and pulsed significances in Figure \ref{ch5_plot15}.
The majority of ensemble members are detected more significantly through
pulsations than through unpulsed emission.  The measured significance for
pulsed detections is a factor of $2.2$ greater than for DC detection.  If we
assume that the detection flux threshold is linear in $\sigma_{DC}$ as it is in
the pulsed significance, this means we require sources to be about twice as
bright on average to detect them through DC emission rather than pulsed
emission, though the factor for a particular pulsar depends strongly on the
light curve morphology, especially the sharpness of the peak(s).

This result also suggests a computationally efficient method for calculating
pulsed flux upper limits.  The average ``efficiency'' factor (2.2 above) can be
determined for some family of light curve templates, and pulsed flux upper
limits can be determined directly from DC upper limits, which may themselves be
straightforwardly calculated from the data with maximum likelihood methods.

Finally, we note that $H_{20w}$ calculated with weights obtained from a spectral
fit using only one degree of freedom---the flux density---is comparable to that
obtained when calculated with both the flux density and the photon index allowed
to vary.  This suggests an insensitivity to the precise spectral shape assumed
for calculation of the weights.  This result has an important practical
implication since the uncertainties on power law parameters can be quite large
for dim sources \citep{1fgl}, in which case assuming a fixed, canonical spectral
shape is likely to yield better results than fitting both flux density and power
law slope.

\section{Discussion}

We have demonstrated that incorporating probability weights into existing
pulsation tests appreciably increases their sensitivity and robustness.  This
increase has nontrivial implications for the detected pulsar population.  E.g.,
the population of pulsars detected so far is observed to have a slope of about
$-1$ on a Log N-Log S plot \citep{psr_cat}, meaning that the increase in
sensitivity translates directly to an increase in the detected population.  That
is, the adoption of weighted statistics can increase the number of pulsars
detected by \fermi by 50--100\%.  The technique will work even better relative
to unweighted tests in the central regions of the Galaxy, within $1\dg$ of the
Galactic plane and within $90\dg$ longitude of the Galactic center, where the
projected source density is high, leading to ample discovery opportunity but
significant source confusion.  These back-of-the-envelope arguments suggest
\fermi may be able to detect, by the end of its mission, the full population of
radio-timed pulsars whose $\gamma$-ray beams intersect the earth.

Increasing the population size is important for attempts to understand the
radiation mechanism of pulsars.  The light curves predicted by models of
magnetospheric emission depend strongly on the configuration of the magnetic
field (represented by the angle between the magnetic dipole and the neutron star
spin axis) and the viewing angle, the angle between the spin axis and the line
of sight.  Although these quantities can sometimes be estimated by, e.g., radio
polarization measurements \citep{weltevrede2010} or X-ray observation of
symmetrical pulsar wind nebulae \citep{ng_and_romani1994}, they are more often
nuisance parameters.  By analyzing a large population simultaneously, one
samples many realizations of the geometry and effectively marginalizes these
nuisance parameters, allowing robust inference of the model parameters.
Although pulsars detected by increasing the sensitivity will be certainly be
faint, even the identification of crude features such as the number of peaks and
their separation can significantly constrain the allowed geometry for the pulsar
\citep{watters_et_al2009}.

Besides improving prospects for detecting the \emph{known} pulsar population,
the machinery developed here to assess statistical performance can be useful for
population synthesis, i.e., placing the strongest constraints possible on the
$\gamma$-ray flux from radio-loud pulsars.  In the material above, we saw that
weighted statistics outperform both unweighted versions of the same statistics
\emph{and} unpulsed significance tests.  Thus, pulsed sensitivity may be the
best tool for constraining the $\gamma$-ray emission.  Although much work has
been done on extracting analytic pulsed upper limits \citep[e.g.]{dejager_3},
the weighted method is not particularly amenable to this approach.  Each source
comes with its own set of probability weights that depends strongly on the
pulsar spectrum and its position on the sky.  With this complication, it makes
sense to instead explore the sensitivity as a function of light curve shape (and
perhaps spectrum) using an ensemble of simulated sources as we have done here.
Although we have only considered a small number of positions, this exercise can
in principle be scaled to a sufficiently fine tessellation of the sky to provide
a pulsed sensitivity map for the full sky.  Finally, this method can be
specialized to provide pulsed flux upper limits for particular known pulsars by
combining simulated data for a candidate pulsar with the data observed by the
\fermin.

\section{Conclusion}
We have shown that existing tests for pulsation gain greatly in sensitivity
through incorporation of probability weights.  The new versions retain their
asymptotic calibration, are insensitive to how data are selected, and allow
detection of pulsars fainter by a factor of 1.5 to 2.  Taken together, these
results represent a significant gain in the search for pulsed $\gamma$ rays
from pulsars and suggest the adoption of weighted statistics---in particular,
$H_{mw}$---for omnibus pulsation tests.

\acknowledgements
The author wishes to thank fellow members of the LAT Collaboration for helpful
discussions and the anonymous referee for suggestions that appreciably improved
the paper.  Support for this work was provided by the National Aeronautics and
Space Administration through Einstein Postdoctoral Fellowship Award Number
PF0-110073 issued by the Chandra X-ray Observatory Center, which is operated by
the Smithsonian Astrophysical Observatory for and on behalf of the National
Aeronautics Space Administration under contract NAS8-03060.

\bibliographystyle{apj}
\bibliography{ms_jan2011}

\appendix
\twocolumngrid

\section{The Asymptotic Null Distribution of $H_m$}
\label{appB}

Recall the $H_m$ statistic is defined as
\begin{equation}
H_m = \mathrm{max}[Z^2_i - c(i-1)]\equiv\mathrm{max}[X_i],\, 1\leq i \leq m.
\end{equation}
In its original formulation, $m=20$ and $c=4>0$ suppresses contributions from
the higher harmonics in the null case.  Let $X_i \equiv Z^2_i - c(i-1)$.  (For
convenience, $X_0 \equiv c$.) Then $H_m$ is an extreme order statistic of the
$X_i$, a collection of $m$ dependent, non-identically distributed rvs.

\subsection{The Joint Probability Density Function of $\vec{X}$}

Assuming $Z^2_m\sim\chi^2_{2m}$, $X_{i+1}$ can be
obtained from $X_{i}$ by adding a $\chi^2_2$ distributed variable and
subtracting $c$:
\begin{equation}
\label{conditional}
f_{X_{i+1}|X_i}(x_{i+1} | x_i) = \chi^2_2(x_{i+1} - x_{i} + c).
\end{equation}

Let $\vec{X}_m$ be a random vector in $\mathbb{R}^m$, such that $H_m$ is the maximum element of $\vec{X}_m$.  We construct the joint pdf for $\vec{X}$ as a product of conditional distributions:
\begin{eqnarray}
f_{\vec{X}_m}(\vec{x}_m) &=& 
f_{X_m|\vec{X}_{m-1}}(x_m |\vec{x}_{m-1})\nonumber\\
&\times&f_{X_{m-1}|\vec{X}_{m-2}}(x_{m-1} | \vec{x}_{m-2})\times\cdots\nonumber\\
&\times& f_{X_2|X_1}(x_2 | x_1)\times f_{X_1}(x_1)\\
&=& \prod_{i=1}^m \chi^2_2(x_i - x_{i-1} + c),
\end{eqnarray}
demonstrating the Markov property.

Inserting the explicit form for $\chi^2_2(x)=\frac{1}{2}\exp(-\frac{x}{2})\,\theta(x)$, where $\theta(x)$ is the Heaviside step function restricting support to positive arguments, yields a significant simplification:
\begin{equation}
\label{explicit}
f_{\vec{X}_m}(\vec{x}_m) = \frac{\alpha^{m-1}}{2}\times\left[\prod_{i=1}^m \theta(x_i - x_{i-1} + c)\right]\exp\left(-\frac{x_i}{2}\right),
\end{equation}
where for convenience $\alpha\equiv \frac{1}{2}\exp\left(-\frac{c}{2}\right)$.

\subsection{The Cumulative Distribution Function of $H_m$}

$H_m$ is just the maximum element of the vector $\vec{X}_m$.  Thus, the
probability to observe a value less than or equal to $h_m$ is simply the
integral of the $f_{\vec{X}}(\vec{x})$ over all values of $\vec{x}$ with all
elements of $\vec{x}$ less than or equal to $h_m$:
\begin{equation}
F_{H_m}(h_m) = \prod_{i=1}^{m}\left(\int_{-\infty}^{h_m} dx_i\right) f_{\vec{X_m}}(\vec{x_m}).
\end{equation}
With the form obtained in Eq. \ref{explicit} the rhs becomes

\begin{equation}
\label{f1}
\frac{\alpha^{m-1}}{2} \prod_{i=1}^{m}\left[\int_{-\infty}^{h_m} dx_i \theta\left(x_i - x_{i-1} + c\right)\right]\exp\left(-\frac{x_i}{2}\right).
\end{equation}

While the integrand is simply an exponential in a single variable, the main
difficulty in evaluating the expression lies in determining the support of the
integrand.

\subsubsection{Reduction of $F_{H_m}(h_m)$}
We can develop the integral in Eq. \ref{f1} recursively.  First, we make a
change of variables in the rightmost integral: $u_m \equiv x_m - x_{m-1} + c$.
This integral is then
\begin{eqnarray}
&&\int_{0}^{h - x_{m-1} + c}du_m\, \exp\left(\frac{-u_m -x_{m-1}+c}{2}\right) =
\nonumber\\&&\alpha^{-1}\left[\exp\left(\frac{-x_{m-1}}{2}\right) - \exp\left(-\frac{h+c}{2}\right)\right].
\end{eqnarray}

The first term of the lhs is the cumulative distribution function for an $m-1$
harmonic H-test, $F_{H_{m-1}}$, while the second term is a volume:
\begin{equation}
\label{f2}
F_{H_m}(h_m) = F_{H_{m-1}}(h_{m}) - \alpha^{m-1}\times\exp\left(-\frac{h_m}{2}\right) I_{m-1},
\end{equation}
where
$$I_n(h) = \prod_{i=1}^{n}\left[\int_{-\infty}^{h} dx_i\, \theta\left(x_i - x_{i-1} + c\right)\right].$$
Fully reducing $F_{H_m}$ yields a power series in $\alpha$:
\begin{equation}
\label{f3}
F_{H_m}(h_m) = 1 - \exp\left(-\frac{h_m}{2}\right)\times\sum_{n=0}^{m-1} \alpha^n I_{n}(h_m).
\end{equation}

\subsubsection{Evaluation of $I_n$}

We begin with a change of variables to eliminate the step functions.  Let $u_i\equiv x_i -\sum_{j=1}^{i-1} (x_j - c)$.  Then
\begin{equation}
I_n(h) = \prod_{i=1}^{n}\left[\int_{0}^{B_i} du_i \right].
\end{equation}
Here, $B_i = h + (i-1)c - \sum_{j=1}^{i-1} x_j$, and we note that $B_n = B_{n-1} + c - u_{n-1}$.  We can evaluate the $n$ integrals recursively.  With each integration, we make the change of integration variable to $q_i \equiv B_{i} + (n-i+1)c - u_i$.  For instance, evaluating the rightmost integral, we have
\begin{eqnarray}
I_n(h) &=& \prod_{i=1}^{n-1}\left(\int_{0}^{B_i} du_i \right) B_n
\\	   &=& \prod_{i=1}^{n-2}\left(\int_{0}^{B_i} du_i \right) \int_0^{B_{n-1}} du_{n-1}\, B_{n-1} + c - u_{n-1}
\nonumber\\     &=& \prod_{i=1}^{n-2}\left(\int_{0}^{B_i} du_i \right) \int_{2c}^{B_{n-1}+2c} dq_{n-1}\, (q_{n-1} - c)
\nonumber\\     &=& \prod_{i=1}^{n-2}\left(\int_{0}^{B_i} du_i \right) \int_{2c}^{B_{n-1}+2c} d\frac{q_{n-1}^2}{2} - cI_{n-1}.
\nonumber
\end{eqnarray}
This form is typical as one continues to integrate.  The change of variable always produces a monomial in the integration variable.  The upper boundary then produces a term in the integration variable of the next integral while the lower boundary produces a monomial of $c$.  This separation allows a recursive development for $I_n$, and combining the recursive terms yields
\begin{equation}
\label{ifinal}
I_n(h) = \frac{(h+nc)^n}{n!} - \sum_{j=1}^{n}I_{n-j-1}\ \frac{(jc)^{j}}{j!}.
\end{equation}
We note that $I_0(h)=1$ and $I_1(h)=h$.

\subsection{Monte Carlo Validation} 
\begin{figure}
\center
\includegraphics[width=3.5in]{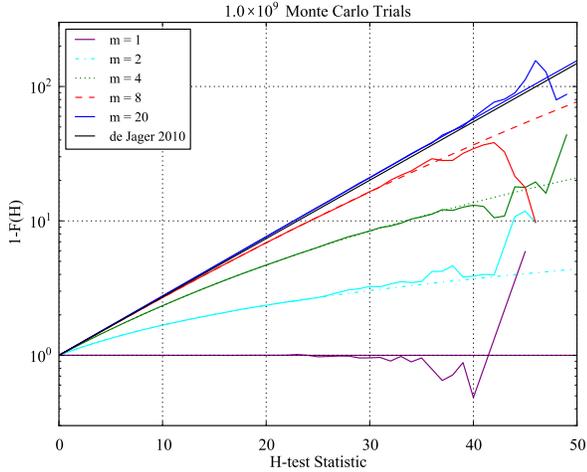}
\caption{The survival function ($1-F(H)$) of the asymptotic distribution for H for a sample ($n=10^{9}$) drawn from the null distribution by simulation.  To reduce the scale, we divide by the survival function for a $\chi^2$ variable with two degrees of freedom, $\exp(-0.5x)$.  For each maximum harmonic, the sample distributions agree with the asymptotic calibration.}
\label{ch5_plot6}
\end{figure}

We validated our results with Monte
Carlo simulations of the $H_{20}$ statistic in the asymptotic null case.
Specifically, for each realization of $H_{20}$, we drew $20$ realizations from
a $\chi^2$ distribution with one degree of freedom and determined $H$
accordingly (with $c=4$.)  In Figure \ref{ch5_plot6}, we show the results of
$10^9$ Monte Carlo trials for a variety of maximum harmonics.  The results are
in good agreement with the asymptotic distribution derived here.

\subsection{General Properties for $m\geq10$}
Although the asymptotic calibration for $H_{20}$ has been previously
characterized by Monte Carlo \citep{dejager_1,dejager_2}, the analytic solution
extends the calibration to an arbitrary collection of harmonics.  (The method
used here to develop the asymptotic distribution can be easily extended to
cases where the harmonics are not sequential, e.g. a test incorporating only
the 2nd, 10th, and 20th harmonics.  However, for simplicity's sake we retain
the original formulation of an inclusive set of harmonics.)

\begin{figure}
\center
\includegraphics[width=3.5in]{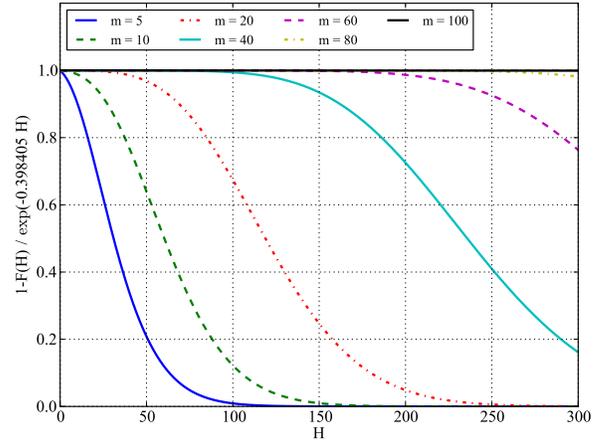}
\caption{The survival function for the $H$-test for a variety of maximum harmonics.  The values have been scaled by $\exp(-0.398405\,H_{m})$, which is seen to provide an excellent approximation to the survival function for large $m$.}
\label{ch5_plot18}
\end{figure}

It is apparent from Figure \ref{ch5_plot6} that the distributions approach a
limiting distribution as $m$, the maximum harmonic, becomes large.  Indeed, as
we see in Figure \ref{ch5_plot18}, for $m\geq10$, there is very little
difference in the tail probability for ``practical'' values of $H$.  For
instance, the chance probabilities of observing $H_m>50$ for any harmonic
$m\geq10$ are all within a factor of 2 of each other, a negligible distinction
at this significance level, and for $H_m<50$, the discrepancy is even smaller.
We conclude that, once one has made the decision to choose $m$ large
enough to allow for sharply-peaked light curves---and indeed, this is the whole
point of the $H$-test---that there is no penalty for making $m$ as large as
feasible.  In this sense, \emph{the $H$-test becomes truly omnibus}, as $m$ can
be chosen large enough to make the test sensitive to light curves with
arbitrarily sharp features.  Finally, we note a practical formula for the cdf
applicable for $m\geq10$: $1-F(H)\approx\exp(-0.398405\,H)$.  This is,
naturally, quite close to the formula reported by \citet{dejager_2}.  For
$m<10$, quadratic terms in the exponential are important for an accurate
evaluation of the tail probability and the full expression Eq. \ref{f3} should
be used.

\end{document}